\documentclass[10pt,journal,compsoc]{IEEEtran}

\ifCLASSOPTIONcompsoc
  \usepackage[nocompress]{cite}
\else
  \usepackage{cite}
\fi

\ifCLASSINFOpdf
\else
\fi
\begin{document}
%
\title{Feature Engineering Using File Layout for Malware Detection}
%
%
%
%

\author{
Jeongwoo Kim,
Eun-Sun Cho,~\IEEEmembership{Member,~IEEE,}
Joon-Young Paik
\IEEEcompsocitemizethanks{
\IEEEcompsocthanksitem Jeongwoo Kim and Eun-Sun Cho were with the Department of Computer Science and Engineering, Chungnam National University, Daejeon, South Korea.
\protect\\
\IEEEcompsocthanksitem J.k Paik was with School of Software, Tiangong University, Tianjin, China.}

\thanks{This manuscript was presented in the poster session of The Annual Computer Security Applications Conference (ACSAC) 2020.}
}

\IEEEtitleabstractindextext{%
\begin{abstract}
Malware detection on binary executables provides a high availability to even binaries which are not disassembled or decompiled. However, a binary-level approach could cause ambiguity problems. In this paper, we propose a new feature engineering technique that use minimal knowledge about the internal layout on a binary. The proposed feature avoids the ambiguity problems by integrating the information about the layout with structural entropy. The experimental results show that our feature improves accuracy and F1-score by 3.3\% and 0.07, respectively, on a CNN based malware detector with realistic benign and malicious samples. 
\end{abstract}

}

\maketitle

\IEEEdisplaynontitleabstractindextext

%
\IEEEpeerreviewmaketitle

\section{Introduction}

%
%
%
%
\IEEEPARstart{C}{NN}(Convolution Neural Network)-based malware detection have been highlighted, recently. CNNs have been known to have excellent performance in pattern recognitions, especially for shift-invariant patterns. The adoption of CNNs makes malware detectors resilient to obfuscation techniques (e.g., code transposition, subroutine reordering) that malware authors have used because obfuscated malware tend to include shift-invariant patterns. Nevertheless, CNN-based malware detection suffers from ambiguity on binary [1]. Binary-level detection deals with a binary as a byte stream. Thus, it is hard to differentiate same or similar patterns that have different meanings.

A structural entropy based feature is one of popular features for malware detection [2-4]. It is represented as a kind of an entropy stream. It is obtained by dividing a binary into chunks and calculating entropy values of every chunk. A structural entropy based feature on a CNN model yields competitive performance for malware classification. However, ambiguity problem still exists in structural entropy because some patterns of entropy values look same or similar but could indicate different functionalities (i.e., metadata, code, icons, images, and so on). 

In this paper, we propose a feature engineering technique to solve the ambiguity problems of structural entropy by leveraging minimal knowledge on the layout (i.e., sections) of a binary. In the proposed feature, information about Section table is combined with structural entropy. It allows to differentiate same or similar entropy patterns in different sections. Then, the new feature is fed into our CNN-based malware detector. Our feature outperforms a feature of an entropy stream by 0.07 of F1-score and 3.4\% of accuracy. 

The rest of this paper is organized as follows. We propose our feature engineering technique and CNN-based malware detector (Section 2). Next, we show the experimental results with realistic samples (Section 3). Finally, we conclude the paper (Section 4).

\section{Basic Idea}
Our feature engineering aims at resolving the ambiguity incurred during malware detection by adding information about the sections in a binary to structural entropy. 

The motivation behind the use of information on sections is that each section has a unique functionality. The same or similar patterns in different sections have distinct semantics, which helps to avoid the ambiguity of patterns. The information on sections which we require is very limited, which is restricted to only the Section tables in the File Header. The minimal information is available from even a binary which is not disassembled or decompiled to higher-level representations for further analyses.

In our feature engineering, 13 sections are differentiated according to the specification of PE file [5]. They are Header, .data, .edata, .idata, .pdata, .rdata, .rsrc, .reloc, .text, .tls, .sdata, .xdata, Undefined. The Header includes DOS MZ header, DOS stub, NT header, Section table and so on. The following 11 sections are specified in the PE specification. The final Undefined indicates any section that is not identified by the other 12 sections. Each section is represented as a one-hot vector, in which a single element in a vector is 1 and the others are 0. 

Our feature engineering consists of the following 7 steps. 

\flushleft
\begin{enumerate}
\item Identify sections in a binary based on the above 13 sections.
   
\item Extract bytes from a binary by the identified sections.
   
\item Divide bytes of individual sections into same sized chunks \begin{math} Chunks^S_i\end{math} means a chunk of the index i in Section S where 0 $\leq$ i $<$ m, and  S $\in$ \{ Header, .data, .edata, .idata, .pdata, .rdata, .rsrc, .reloc, .text, .tls, .sdata, .xdata, Undefined \}.

\item Calculate entropy values of every chunk of the binary. \begin{math} Entropy^S_i\end{math} is an entropy of a chunk of the index i in Section S.

\item Make one-hot vectors of every chunks. \begin{math} OhV^S_i\end{math} is a one-hot vector of a chunk of the index i in Section S. For instance, the 10th chunk in .edata is represented as \begin{math} OhV^.edata_10\end{math}, and its value is $<$0,0,1,0,0,0,0,0,0,0,0,0,0$>$ when the 3rd element of the vector corresponds to .edata section

\item Combine \begin{math} Entropy^S_i\end{math} and \begin{math}OhV^S_i\end{math}. The resultant vector is 
called \begin{math}chunkVec^S_i\end{math}. For instance, when \begin{math} Entropy^.edata_10\end{math} is 2.3, \begin{math}chunkVec^.edata_10\end{math} is $<$2.3,0,0,1,0,0,0,0,0,0,0,0,0,0$>$ with the above \begin{math}OhV^.edata_10\end{math}.

\item Concatenate every chunkVec of the sections in the binary. Therefore, a feature of a binary becomes a two-dimensional vector that has a shape of <m,14>. 
   
\end{enumerate}

After the feature engineering phase, the features are fed into the malware detection model. 

In the model, the three convolution layers, each of which is followed by a pooling layer, extract high-level features from our input feature. For this, 1D convolution filters move along the features; a filter \begin{math} w \in R^{l\times 14} \end{math} where l is the size of a filter. The three fully connected layers and output layer follow the convolution layers. We adopt a Rectified Linear Unit (ReLU) function as an activation function. Every layer is normalized by BatchNorm1D. 

\section{Experiments}
For the experiments, we collected normal PE files and malicious PE files respectively. Benign files (i.e., extension .exe and .dll) were collected from System32 folders in Windows and malicious PE files were collected from VirusShare\_0 [6]. The detail on the data sets, Benign and Malware, is shown in Table 1. First, each of Benign and Malware sets was split into a train set and a test set at a ratio of 70/30. Next, we conducted our feature engineering technique to all the datasets while we set the chunk size to 4096 bytes. Finally, the features were fed into the CNN-based malware detection model. To handle variable-length input in our CNN model, the length of an input was limited to 3,600 by padding zero vectors to it or truncating it. Thus, an input feature of a file was represented as a vector of <3600, 14>.

\begin{table}[!h]
\centering
\caption{Detail of used dataset}
\begin{tabular}{lll}
\hline
\hline
Family         & ClassID & Total  \\ \hline
Benign         & 0       & 4,870  \\
Malware        & 1       & 16,712 \\ \hline
\hline
\end{tabular}
\end{table}

We evaluated how important the information about the sections of PE file is for malware detection. For this, we compared the proposed feature with a different feature, entropy streams of chunks of binaries. Our feature and the entropy streams were extracted from Benign and Malware sets. The features were fed into CNN-based malware detector for training and testing. Our feature improved the accuracy by 3.3\% and the macro-averaged F1-score by 0.05, compared with the entropy stream, as shown in Table 2. It means that the information on the sections relieves the ambiguity problems, and the one-hot vectors of the chunks about the sections could make informative patterns by themselves.

\begin{table}[!h]
\centering
\caption{Performance comparison}
\begin{tabular}{lll}
\hline
\hline
Feature         & Accuracy & F1-score  \\ \hline
Entropy streams \\(w/o information on sections)         & 95.8\%       & 0.94  \\
The Proposed one\\ (w/ information on sections)         & 99.1\%       & 0.99  \\ \hline
\hline
\end{tabular}
\end{table}

\section{Conclusion}
In this paper, our feature engineering technique solves the ambiguity problems by integrating the information on sections with the entropy stream of the chunks in a binary. The proposed feature showed better accuracy and F1-score with realistic samples than the feature of an entropy stream. We figure out that the information about the sections of PE file can improve performance of malware detection in this experiment. We plan to further analyze the Undefined section in order to detect malware which exploit various packers, crypters, and protectors.

\section{Acknowledgment}
This work was supported by Institute for Information \& communications Technology Planning \& Evaluation(IITP) grant funded by the Korea government(MSIT) (No.2019-0-01343, Regional strategic industry convergence security core talent training business), in part by the National Natural Science Foundation of China (NSFC) under Grant 61806142 and by the Natural Science Foundation of Tianjin under Grant 18JCYBJC44000, by the Tianjin Science and Technology Program under Grant19PTZWHZ00020.

\end{document}